\documentclass{ws-procs975x65--arXiv}

\begin{document}

\title{APPARENT ACCELERATION AND AN ALTERNATIVE CONCORDANCE FROM CAUSAL BACKREACTION}

\author{BRETT BOCHNER}

\address{Department of Physics and Astronomy, Hofstra University,\\
Hempstead, NY 11549, USA\\
E-mails: brett\_bochner@alum.mit.edu, phybdb@hofstra.edu}

\begin{abstract}
A phenomenological formalism is presented in which the apparent acceleration 
of the universe is generated by cosmic structure formation, without resort 
to Dark Energy, modifications to gravity, or a local void. The observed acceleration 
results from the combined effect of innumerable local perturbations due to 
individually virializing systems, overlapping together in a smoothly-inhomogeneous 
adjustment of the FRW metric, in a process governed by the causal flow of 
inhomogeneity information outward from each clumped system. After noting how 
common arguments claiming to limit backreaction are physically unrealistic, 
models are presented which fit the supernova luminosity distance data essentially 
as well as $\Lambda$CDM, while bringing several important cosmological parameters 
to a new Concordance. These goals are all achieved with a second-generation 
version of our formalism that accounts for the negative feedback of 
Causal Backreaction upon itself due to the slowed propagation of gravitational 
inhomogeneity information.
\end{abstract}


\bodymatter

\section*{Introduction and Overview for Causal Backreaction}\label{bdb:sec1}

Evidence that the expansion rate of the universe is accelerating 
\cite{RiessAccel98,PerlAccel99} has led to the conclusion that the 
dominant cosmic component is ``Dark Energy'', a negative pressure 
(yet smoothly-distributed) material. But given problems related to 
magnitude fine-tuning \cite{KolbTurner} and ``coincidences'' 
\cite{ArkaniHamedCoinc} for a Cosmological Constant ($\Lambda$) -- 
and more unknowns for evolving forms of Dark Energy -- 
an attractive alternative is known as ``backreaction'', 
in which the (apparent) acceleration is a natural result of the 
ever-increasing inhomogeneity of the structure-forming universe 
(see, e.g., Refs.~\citen{Bochner21Texas,SchwarzFriedFail,RasanenFriedFail,
KMNRinhomogExp,WiltInhomogNoDE}). 

Presently, backreaction is not widely considered to be a viable method 
for achieving the observed acceleration \cite{SchwarzBackReactNotYet}. 
Elsewhere \cite{BochnerAccelPapers0I}, I discuss how this seeming 
inadequacy of backreaction is likely due to the neglect of crucial physics 
via the use of overly trivial backreaction models. Some neglected 
physics includes: the dropping of vorticity for the virialization of 
stabilized structures; the dropping of ``small amplitude'' (but 
{\it cumulatively} important) terms in perturbation theory such as 
time derivatives, and velocity terms at least up to $O[(v/c)^2]$; 
the neglect of tensor components and `gravitomagnetic' terms, which 
carry gravitational perturbation information {\it causally}, at $c$; 
the neglect of the cumulative {\it overlap} of different perturbations, 
artificially isolating each one from all others (they just ``fall 
out of the expansion''), as in Swiss-Cheese models; and the debatable 
view (from Newtonian cosmology) that (individually) Newtonian-strength 
perturbations must yield essentially no backreaction.

Our approach here, in contrast, is to try to include {\it all} of the 
necessary physics of backreaction -- even at the cost of using a
very simplified model -- rather than using a more sophisticated 
(or even exact) model of a {\it physically} simplified universe.

\begin{figure}
\begin{center}
\psfig{file=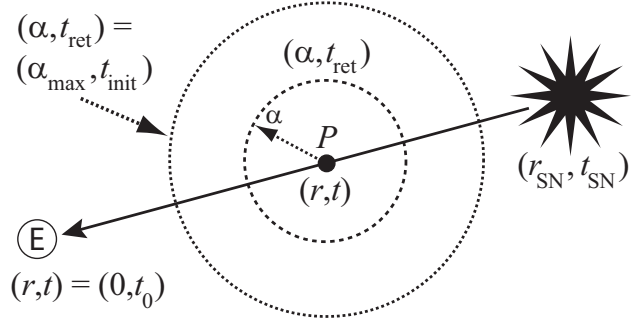,width=3.25in}
\caption{Geometry for computing the inhomogeneity-perturbed 
metric at each point along the integrated path of a light ray 
from a supernova to our observation point at Earth.}
\label{FigSNRayTraceInts}
\end{center}
\end{figure}

Utilizing the straightforward nature of the ``before'' vs. ``after'' 
cosmic states -- that is, quantifying the total effect of the 
transition from smooth universe to `fully clustered' -- 
it can be argued \cite{BochnerAccelPapers0I} that the 
net effect of the formation of a vorticity-stabilized, virialized cluster 
far from a given observation point $P$, should be the simple addition of 
the Newtonian metric perturbation due to that cluster's stabilized mass, 
to the overall (initially FRW) metric at $P$. For any point $P$, since 
the number of perturbing contributions increases with distance as $r^2$, 
but each contribution only weakens as $1/r$, the summed effect of all 
such contributions can become dominantly large -- limited only by the 
look-back distance out to which an observer at $P$ can `see' stabilized 
structures, at that given time. The total metric perturbation (at {\it any} 
given observation point, in a ``smoothly-inhomogeneous'' universe) as a 
function of time, $I(t)$, is thus given by an integral over the (retarded) 
clumping that is felt from great distances at the observation point, as per 
Eq.~\ref{EqnItotIntegration} 
(and as depicted in Fig.~\ref{FigSNRayTraceInts}):
\begin{equation}
I(t) = 
\int^{\alpha _{\mathrm{max}} (t, t_\mathrm{init})}_{0}
\{12 ~ 
\Psi [t_{\mathrm{ret}} (t, \alpha)] 
~ [(t_{0} / t)^{2/3}] \} ~ 
\alpha ~ d \alpha ~ .
\label{EqnItotIntegration}
\end{equation}

To an observer at $P$, new clustered masses thus seem to 
`appear' over time (as attractive, perturbing forces) as the 
material within each one transitions from smoothly distributed 
and almost motionless, to clumped and vorticity-stabilized. 
And though each such perturbation is `Newtonian' in strength 
(as seen at $P$), the total, summed metric perturbation 
will grow ever stronger as more virialized clusters 
`come into view', eventually becoming strong enough to generate 
the apparent acceleration. This accumulation of perturbation 
effects from cosmologically-distant new structures, 
coming in towards $P$ at speed $c$, is called 
``Causal Updating'' \cite{BochnerAccelPapers0I}.

Averaging over clumps distributed randomly in direction, 
we get the metric: 
\begin{equation}
ds^{2} = 
- c^{2} [ 1 - I(t) ] ~ dt^{2} 
~ + ~ \{ [a_{\mathrm{MD}}(t)]^{2} ~ 
[ 1 + (1/3) I(t) ] \} 
~ \vert d \vec{r} \vert ^{2} ~ . 
\label{EqnFinalBHpertMetric}
\end{equation}
This is the final metric (computed for any given 
``clumping evolution model'', $\Psi [t]$) to be used for calculating 
Hubble curves, and all cosmological parameters of interest. 

Though several models were found in Ref.~\citen{BochnerAccelPapers0I} 
which provided acceptable fits to the SCP Union1 SNIa data 
\cite{KowalRubinSCPunion}, those initial backreaction simulations neglected 
an important complication: the fact that {\it old} metric perturbations 
from pre-existing structures slows down all {\it future} propagation of 
inhomogeneity information (from new structures, from old structures 
even farther away, etc.). This weakens Causal Updating, so that 
Causal Backreaction has a negative feedback upon itself 
(making an ``eternal'' acceleration very unlikely here). 
The Causal Backreaction response to clustering is therefore `recursive' 
and nonlinear in terms of response versus clustering strength, and thus 
a second-generation model of Causal Backreaction was designed to 
incorporate the effects of these ``Recursive Nonlinearities'' 
(as distinct from {\it gravitational} nonlinearities). This 
new formalism, and a full suite of simulation runs, fit results, 
and cosmological parameters obtained with it, are given in 
Ref.~\citen{BochnerAccelPaperIII}.

\begin{figure}
\begin{center}
\psfig{file=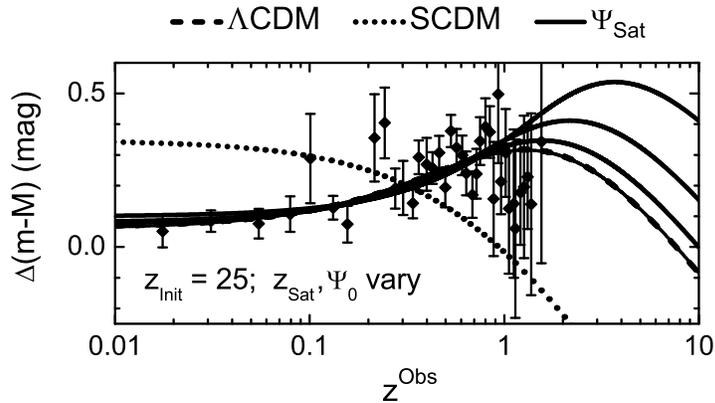,width=4.25in}
\caption{Residual Hubble diagrams for models with different 
``clustering saturation'' epochs $(z_\mathrm{Sat})$, with the 
ultimate degree of clustering $(\Psi _{0})$ optimized for each 
case. Also plotted here are the binned SCP Union1 SNIa data points, 
along with flat SCDM and Concordance $\Lambda{\mathrm{CDM}}$ cosmologies.}
\label{FigVaryingEarlySatZval}
\end{center}
\end{figure}

The principal result of this study, is that astrophysically realistic 
Causal Backreaction models can indeed be chosen which successfully 
mimic the apparent acceleration generically attributed to $\Lambda$, 
but without any form of Dark Energy, Voids, etc. Some of these models, 
plotted against the Union1 SNIa data, are shown in 
Fig.~\ref{FigVaryingEarlySatZval}.

Though successful at reproducing the `acceleration', quite a large 
amount of clustering is needed -- a nearly-complete clustering 
on several different hierarchical scales, simultaneously 
(\dots Stellar Clusters, Galaxies, Galaxy Clusters\dots). 
A more moderate amount of final clustering is sufficient, 
though, for models in which structure formation saturates at late times 
due to ``gastrophysics'' feedback, and due to clustering slow-down 
(still present in this formalism) from the `acceleration' itself. 

From the final results for our `best-fitting' models found 
(even without a rigorous optimization), the output cosmological 
parameters were calculated, as would be seen by cosmic observers 
at $z = 0$ (such as ourselves). These values are presented 
in Table \ref{TabRNLearlySatRunsAbbrev}, where it is clear 
that there are backreaction models for which all 
of the cosmological parameters considered here -- including the 
observed Hubble Constant ($H^\mathrm{Obs}_{0}$), the observed age 
of the universe ($t^\mathrm{Obs}_{0}$), the matter density required 
for spatial flatness ($\Omega^\mathrm{FRW}_\mathrm{M} = 1$), 
the characteristic angular scale of the CMB acoustic peaks 
($l^\mathrm{Obs}_{\mathrm{A}}$), and the strength of the 
apparent acceleration ($w^\mathrm{Obs}_{0}$) -- are each broadly 
consistent with those from the $\Lambda$CDM cosmologies 
of the ``Concordance Model''. It can therefore be concluded that 
a new, Alternative Concordance can indeed be achieved with 
Causal Backreaction, without any mysterious ``Dark Energy'' 
cosmic component being required. Finally, higher-order terms 
in the expansion ($j^\mathrm{Obs}_{0}$, etc.) could potentially 
be used to distinguish between $\Lambda$CDM and Causal Backreaction.

\begin{table}
\tbl{Output Cosmological Parameters from `Best' Runs with Recursive Nonlinearities.}
{\begin{tabular}{@{}cccccccccc@{}}
\toprule
$z_\mathrm{Sat}$ & 
$\Psi _{0, \mathrm{Opt}}$ & 
$\chi^{2}_{\mathrm{Fit}}$ & 
$H^\mathrm{Obs}_{0}$ & 
$H^\mathrm{FRW}_{0}$ & 
$t^\mathrm{Obs}_{0}$ & 
$\Omega^\mathrm{FRW}_\mathrm{M}$ &
$w^\mathrm{Obs}_{0}$ & 
$j^\mathrm{Obs}_{0}$ & 
$l^\mathrm{Obs}_{\mathrm{A}}$ 
\\
\colrule
\multicolumn{10}{c}{{\it Causal Backreaction simulation runs, 
with the beginning of clumping at $z_\mathrm{init} = 25$}}\\ 
\colrule
0 & 4.1 & 311.8 & 70.07 & 42.32 & 
13.64 & 0.943 & -0.751 & 1.73 & 294.5 \\
0.25 & 2.6 & 313.5 & 69.60 & 40.24 & 
14.00 & 1.054 & -0.620 & 0.15 & 289.7 \\
0.5 & 2.3 & 316.6 & 69.40 & 36.32 & 
14.65 & 1.338 & -0.585 & -0.14 & 279.8 \\
1 & 2.2 & 320.2 & 68.77 & 29.54 & 
15.75 & 2.086 & -0.488 & -0.94 & 259.9 \\
\colrule
\multicolumn{10}{c}{{\it Comparison Values from 
the Union1-best-fit flat $\Lambda{\mathrm{CDM}}$ model 
$(\Omega _{\Lambda} = 0.713 = 1 - \Omega _\mathrm{M})$}} 
\\ 
\colrule
--- & --- & 311.9 & 69.96 & 69.96 & 
13.64 & 0.287 & -0.713 & 1.0 & 285.4 \\
\botrule
\end{tabular}
}
\label{TabRNLearlySatRunsAbbrev}
\end{table}

\section*{Acknowledgments}

I am grateful to Hofstra University for their support of 
my attendance at MG13.

\end{document}